\documentclass[12pt,english,aps,preprint,groupedaddress,superscriptaddress]{revtex4}
\usepackage[T1]{fontenc}
\usepackage[latin9]{inputenc}
\usepackage{color}
\usepackage{textcomp}
\usepackage{amstext}
\usepackage{graphicx}

\makeatletter

\providecommand{\tabularnewline}{\\}

\@ifundefined{textcolor}{}
{%
 \definecolor{BLACK}{gray}{0}
 \definecolor{WHITE}{gray}{1}
 \definecolor{RED}{rgb}{1,0,0}
 \definecolor{GREEN}{rgb}{0,1,0}
 \definecolor{BLUE}{rgb}{0,0,1}
 \definecolor{CYAN}{cmyk}{1,0,0,0}
 \definecolor{MAGENTA}{cmyk}{0,1,0,0}
 \definecolor{YELLOW}{cmyk}{0,0,1,0}
}

\makeatother

\usepackage{babel}
\begin{document}
\author{Hai-Yuan Cao}
\affiliation{State Key Laboratory of Surface Physics, Key Laboratory for Computational Physical Sciences (MOE), Department of Physics, Fudan University, Shanghai 200433, China}

\author{Shiyong Tan}
\affiliation{State Key Laboratory of Surface Physics, Key Laboratory for Computational Physical Sciences (MOE), Department of Physics, Fudan University, Shanghai 200433, China}
\affiliation{Advanced Materials Laboratory,Fudan University, Shanghai 200433, China}

\author{Hongjun Xiang}
\affiliation{State Key Laboratory of Surface Physics, Key Laboratory for Computational Physical Sciences (MOE), Department of Physics, Fudan University, Shanghai 200433, China}

\author{D. L. Feng}
\affiliation{State Key Laboratory of Surface Physics, Key Laboratory for Computational Physical Sciences (MOE), Department of Physics, Fudan University, Shanghai 200433, China}
\affiliation{Advanced Materials Laboratory,Fudan University, Shanghai 200433, China}

\author{Xin-Gao Gong}
\affiliation{State Key Laboratory of Surface Physics, Key Laboratory for Computational Physical Sciences (MOE), Department of Physics, Fudan University, Shanghai 200433, China}

\preprint{This line only printed with preprint option}

\title{The interfacial effects on the spin density wave in FeSe/SrTiO$_{3}$
thin film}
\begin{abstract}
Recently, the signs of both superconducting transition temperature
(T$_{c}$) beyond 60 K and spin density wave (SDW) have been observed
in FeSe thin film on SrTiO$_{3}$ (STO) substrate, which suggests
a strong interplay between superconductivity and magnetism. With the
first-principles calculations, we find that the substrate-induced
tensile strain tends to stabilize the SDW state in FeSe thin film
by enhancing of the next-nearest-neighbor superexchange antiferromagnetic
interaction bridged through Se atoms. On the other hand, we find that
when there are oxygen vacancies in the substrate, the significant
charge transfer from the substrate to the first FeSe layer would suppress
the magnetic order there, and thus the high-temperature superconductivity
could occur. In addition, the stability of the SDW is lowered when
FeSe is on a defect-free STO substrate due to the redistribution of
charges among the Fe 3d-orbitals. Our results provide a comprehensive
microscopic explanation for the recent experimental findings, and
build a foundation for the further exploration of the superconductivity
and magnetism in this novel superconducting interface.

\end{abstract}
\maketitle

\section{introduction}

Magnetism seems to be always involved in the superconducting mechanism
of high-T$_{c}$ superconductors. Often, the superconductivity occurs
when the long-range magnetic order is suppressed somehow, and yet
there are underlying spin fluctuations and antiferromagnetic (AFM)
interactions that could mediate the Cooper paring of the electrons
\cite{Hu2012}. Recently, a large superconducting gap in monolayer
FeSe thin film grown on STO substrate was observed by both scanning
tunneling spectroscopy (STS) \cite{Wang2012} and angle-resolved photoemission
spectroscopy (ARPES) \cite{He2013,Tan2013,Liu2012}, with the superconducting
transition likely at 65 K. This would establish a new T$_{c}$ record
for the iron based superconductors. Intriguingly, it is well known
that the bulk FeSe only exhibits a T$_{c}$ around 8 K \cite{Hsu2008,Song2011},
or 37 K under compressional pressure \cite{Medvedev2009}. It is thus
remarkable to observe a higher T$_{c}$ in monolayer FeSe on STO,
which is under the tensile strain imposed by the substrate. 

For FeSe bulk material and thin film, a collinear $2\times1$ SDW
order is theoretically predicted to be the ground state, similar to
that in the iron pnictides \cite{Subedi2008,Bazhirov2013,Liu2012a,Ma2009}.
Tan \textit{et al.} have substantiated the presence of spin density
wave (SDW) in multilayer FeSe thin films which were grown layer by
layer on the STO substrate with molecular beam expitaxy, and they
showed that when the tensile strain decreases as the lattice relaxes
with increasing thickness, the strength of the SDW decreases as well
\cite{Tan2013}. To our knowledge, no previous theoretical study has
been focused on the evolution of the magnetism with the lattice constant
in this system. According to the strong interplay between superconductivity
and magnetism, the study on how the interfacial effect influences
the magnetism in FeSe/STO thin film should be helpful for the understanding
of the superconductivity there.

The SDW would have been the most prominent in monolayer FeSe next
to the STO due to the most pronounced strain from the substrate, had
it not been suppressed by the charge transferred from the oxygen-vacant
substrate, as suggested by both the experiment \cite{Tan2013} and
theory \cite{Bang2013}. However, previous theoretical studies were
not conclusive. Liu $et$ $al.$ analyzed the orbital-resolved partial
density of states (PDOS) from the density functional theory (DFT)
calculations, and did not find substantial charge transfer between
FeSe and the STO substrate \cite{Liu2012a}. On the other hand, Zheng
$et$ $al.$ predicted a considerable charge transfer from surface
O atoms of STO substrate to Fe atoms of FeSe monolayer \cite{Zheng2013}.
Another recent theoretical paper excluded the Se vacancy in FeSe as
the source of electron doping for superconductivity \cite{Berlijn2013},
which further indicates that the oxygen vacancy on the STO substrate
should play indispensable role in the Fe-HTS. For these reasons, it
is necessary to further investigate how the STO substrate and the
oxygen vacancies there influence the magnetic order in the monolayer
FeSe. 

In this paper, based on DFT calculations we reveal how the strain,
the interfacial coupling and the oxygen-vacant STO substrate affect
the magnetism in the FeSe thin film. We find the strain could enhance
the Fe-Se-Fe superexchange by increasing the Fe-Se-Fe bond angle.
With the increasing superexchange interaction, the local AFM exchange
interaction would be enhanced. The interaction between the STO substrate
and the monolayer FeSe would reduce the charge density in the spin-majority
d$_{xz}$/d$_{yz}$-orbital states of the Fe atoms, which suppresses
the effect of superexchange and reduces the stability of SDW in the
monolayer FeSe. If the oxygen vacancies exist on the surface of STO
substrate, a certain amount of charge would be transferred from the
substrate to the spin-minority d$_{xz}$/d$_{yz}$-orbital states
of the Fe atoms in the monolayer FeSe, which would suppress the SDW
and allow the superconductivity to occur. Meanwhile, the original
symmetry of spin configuration in monolayer FeSe would also be disturbed
due to the oxygen vacancies on the substrate. All of our results are
in close agreement with the recent experiment \cite{Tan2013}. It
thus provides a detailed microscopic understanding of the interfacial
effects in this intriguing system. More importantly, our results suggest
that the high T$_{c}$ in the monolayer FeSe is closely related to
the large underlying superexchange interactions caused by its expanded
lattice, which again highlight the pivotal role of magnetism in the
high temperature superconductivity of iron-based superconductors.

\section{methods}

To study how the substrate affects the monolayer FeSe grown on STO
(001) surface, we carry out the spin-polarized first-principles calculations
using the project augmented wave pseudopotential \cite{Blochl1994,Kresse1999}
implemented in the VASP code \cite{Kresse1996,Kresse1996a}. We employ
the generalized gradient approximation (GGA) of Perdew-Burke-Ernzerhof
for the exchange-correlation-potentials \cite{Perdew1996}. The kinetic
energy cutoff of the plane-wave basis is chosen to be 400 eV. The
force on all relaxed atoms after the optimization is smaller than
0.01 eV/\AA . A $6\times6\times1$ k-point mesh \cite{Monkhorst1976}
for the Brillouin zone sampling and a width of 0.1 eV for the Guassian
smearing are adopted. In all the calculations we employ a local Coulomb
repulsion GGA+U approach for Ti 3d electrons with $U_{Ti}=2$ eV \cite{Pavlenko2012,Pavlenko2012a,Breitschaft2010}.
The electric field induced by the asymmetric relaxed STO is compensated
by a dipole correction \cite{Neugebauer1992}. We also use the $J_{1}-J_{2}$
Heisenberg model to describe the magnetic interactions, while the
exchange parameters are fitted to the total energy of the first principle
calculations. 

\begin{figure}
\begin{centering}
\includegraphics[scale=1.0]{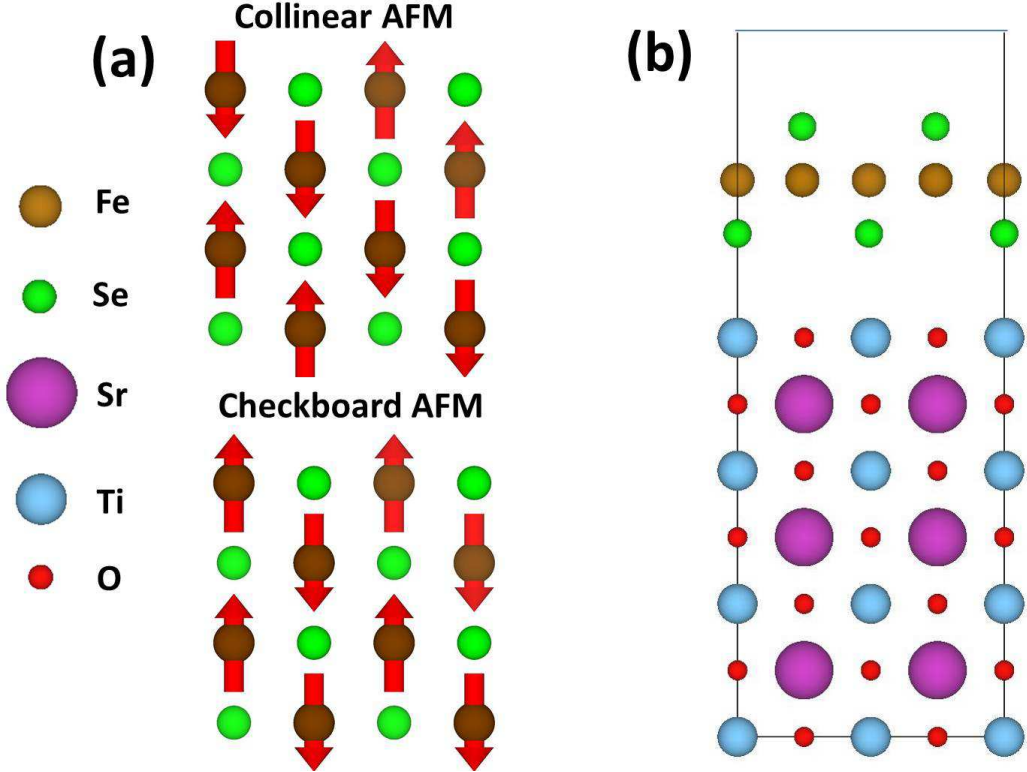}
\par\end{centering}

\caption{ (a) Top view of atomic structures of $2\times2$ supercell monolayer
FeSe and spin patterns of Fe in the AFM state and the SDW state. (b)
Side view of atomic structure of monolayer FeSe on TiO$_{2}$ terminated
STO (001) surface.}
\end{figure}

As shown in Figure 1(a), we use the $2\times2$$\times1$ supercell
to describe the magnetic order of the monolayer and bulk FeSe. To
model the interface with monolayer FeSe on STO, we use a seven-layer
STO (001) slab with the $2\times2$ monolayer FeSe supercell on the
top side plus a vacuum layer about 10 \AA  \ as shown in Figure
1(b). According to previous results \cite{Liu2012a}, the most stable
interfacial configuration is the monolayer FeSe on the TiO$_{2}$
terminated STO surface. In the present calculation, we fix the lattice
constant $a$ to be 3.905 \AA , the lattice constant of the bulk
STO \cite{El-Mellouhi2011}. In the structural optimization, the top
two layers of STO substrate and all FeSe atoms are allowed to relax,
while the atoms in bottom layers of STO substrate are fixed at their
bulk positions. To study the effect of the oxygen vacancies on the
STO substrate, we choose one, two, four vacancies out of eight oxygen
atoms on the $2\times2$ supercell of TiO$_{2}$ terminated surface,
corresponding to $12.5\%,$ $25\%$ and $50\%$ vacancy concentration,
respectively.

\section{results and discussion}

For both free-standing FeSe and epitaxial monolayer FeSe on STO substrate,
we have calculated four magnetic states, including the nonmagnetic
state (NM state), the ferromagnetic state, the checkboard AFM state
(AFM state) and the collinear AFM state (SDW state). The spin texture
of the AFM state and the SDW state in FeSe are shown in Figure 1(a)
\cite{Liu2012a,Bazhirov2013}, respectively. By calculating and comparing
the energy difference between these four states, we find that the
ground state is the SDW state with a large magnetic moment of $\sim2.4$
$\mu_{B}$ on each Fe atom for both free-standing FeSe and epitaxial
FeSe on STO. The ferromagnetic state has very high energy, $\sim0.2$
eV per Fe atom higer than the NM state, so we ignore the ferromagnetic
state in the following discussions. This has also been proved in the
previous theoretical studies \cite{Liu2012a,Ma2009}. 

\begin{figure}
\includegraphics[scale=1.0]{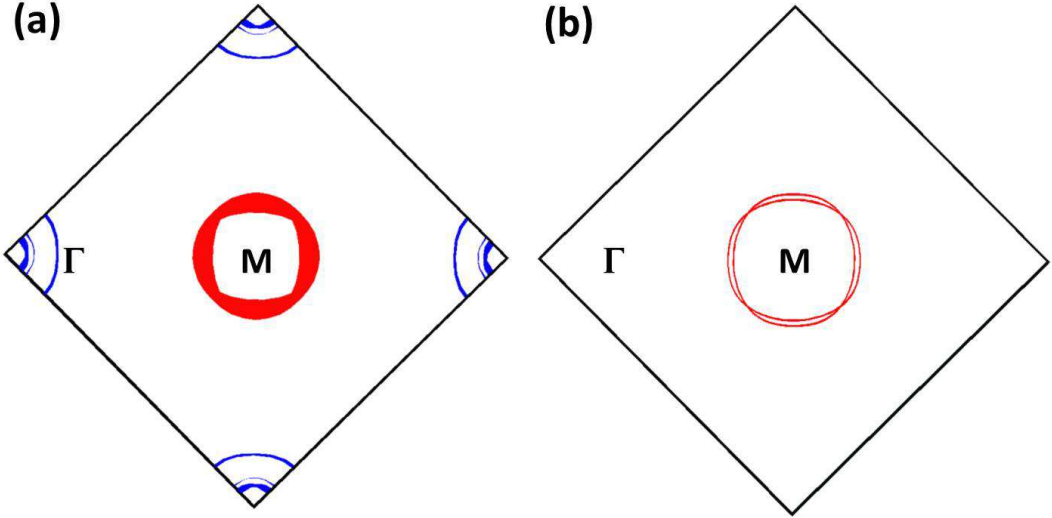}

\caption{\textcolor{black}{The calculated Fermi surface of (a) bulk FeSe and
(b) monolayer FeSe on STO substrate. The center is M point and the
corner is $\Gamma$ point. The electron pockets are denoted as red
while the hole pockets are denoted as blue. The hole pockets are absent
in the Fermi surface of monolayer FeSe on STO substrate\cite{Tan2013,Liu2012,He2013}.}}
\end{figure}

Firstly, we calculate the Fermi surface of bulk FeSe, the obtained
result shown in Figure 2(a) is in close agreement with previous calculations
\cite{Subedi2008} which is composed of several hole pockets around
the $\Gamma$ point and two electron pockets around the M point. For
the Fermi surface of the monolayer FeSe on STO substrate with lattice
constant 3.905 \AA\  , as showing in Figure 2(b), we shift the Fermi
level according to the 0.12 e$^{-}$ per Fe atom as suggested by the
experiment \cite{Tan2013}. It is in good agreement with that observed
in the ARPES experiments \cite{Liu2012,He2013,Tan2013} which is composed
of two nearly degenerate electron pockets around the M point and the
hole pockets around the $\Gamma$ point are absent. These results
confirm that our method can well reproduce the electronic structures
for both bulk FeSe and monolayer FeSe on STO substrate observed in
experiments.

\subsection{How the Tensile Strain Affects the SDW in FeSe}

We firstly study how the external strain could affect the stability
of the SDW state. We use the energy difference between the SDW state
and other two states, the AFM state and the NM state, to assess the
stability of the SDW state \cite{Bazhirov2013,Liu2012a}. In Figure
3, we can see that the energy difference relative to the SDW state
increases with the expanding of lattice constant, which indicates
that the tensile strain can enhance the stability of the SDW state
in both bulk and monolayer FeSe. Moreover, the energy difference in
the monolayer and bulk FeSe with the same lattice constant is almost
the same, therefore the relative stability of the SDW is insensitive
to the thickness of the FeSe, but sensitive to the lattice contant. 

\begin{figure}
\includegraphics[scale=1.0]{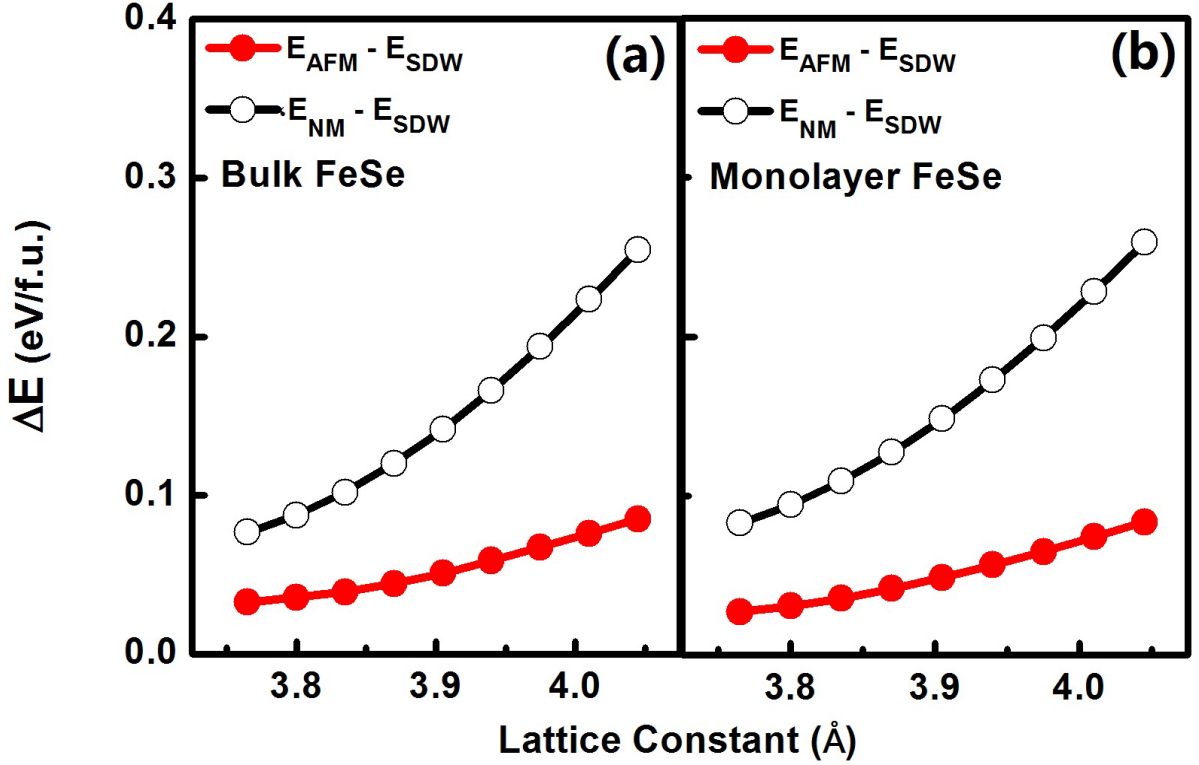}

\caption{The calculated energy difference (relative to the SDW state) of the
AFM state (red circle) and the NM state (black circle) versus the
lattice constant of FeSe, (a) for bulk FeSe and (b) for monolayer
FeSe, respectively. In both the bulk and monolayer FeSe, the SDW state
tends to be energetically more stable with ncreasing lattice constant.}
\end{figure}

To investigate why the tensile strain can affect the stability of
SDW, we would model the magnetic interaction in FeSe with different
lattice constant. We assume that the interaction between the Fe spins
dominates the energy difference between different magnetic orders.
We could map the magnetic interaction to the following Heisenberg
model which is described by the nearest-neighbor and next-nearest-neighbor
coupling parameters $J_{1}$ and $J_{2}$ \cite{Ma2009}:
\begin{equation}
H=J_{1}\sum_{\langle ij\rangle}\vec{S_{i}}\cdot\vec{S_{j}}+J_{2}\sum_{\langle\langle ij\rangle\rangle}\vec{S_{i}}\cdot\vec{S_{j}}
\end{equation}
whereas $\langle ij\rangle$, $\langle\langle ij\rangle\rangle$ denote
the summation over the nearest-neighbors and the next-nearest-neighbors,
respectively. Using the method proposed by previous theoretical work\cite{Ma2008a},
we can determine the value of $J_{1}$ and $J_{2}$. In bulk FeSe,
we find $J_{1}=74$ meV/S$^{2}$ and $J_{2}=43$ meV/S$^{2}$, in
close agreement with the previous results\cite{Ma2009}. 

\begin{table}
\begin{centering}
\begin{tabular}{|c|c|c|c|c|}
\hline 
Lattice Constant ($\text{\AA}$) & J$_{1}$ (meV) & J$_{2}$ (meV) & J$_{2}$/J$_{1}$ & Fe-Se-Fe Angle\tabularnewline
\hline 
\hline 
3.765 & 74 & 43 & 0.58 & 105\textdegree{}\tabularnewline
\hline 
3.905 & 78 & 53 & 0.68 & 109\textdegree{}\tabularnewline
\hline 
4.045 & 82 & 60 & 0.73 & 113.5\textdegree{}\tabularnewline
\hline 
\end{tabular}
\par\end{centering}

\caption{Structural parameters, calculated nearest-neighbor and next-nearest-neighbor
coupling parameters of monolayer FeSe. The definition of Fe-Se-Fe
angle is shown in Figure 4(b). Here we assume S = 1 for Fe atoms.}
 
\end{table}

As shown in Table 1, $J_{1}$ increases only slightly with increasing
lattice constant, but $J_{2}$ increases about 40\% when the lattice
constant increases just a few percent. More clearly, $J_{2}/J_{1}$
increases monotonously with lattice constant expanding, suggesting
that the SDW state is getting more and more stable. According to the
frustrated Heisenberg model, the magnetic exchange energy in the unit
cell of FeSe with four Fe atoms is $-2J_{1}+2J_{2}$ for the AFM state
and $-2J_{2}$ for the SDW state (assuming S = 1). It is known that
the frustration between $J_{1}$ and $J_{2}$ destructs the stability
of the AFM state and induces the SDW state when $J_{2}>J_{1}/2$ for
a square lattice \cite{Ma2009}. 

To explore the origin of the changing of $J_{1}$ and $J_{2}$ with
the lattice constant, we have calculated the charge distribution around
Fe and Se atoms. As shown in Figure 4(a) and 4(b), there is almost
no charge density between two nearest-neighbor Fe atoms but the bonds
are formed between Fe and Se atoms. Similar to the previous calculation
of LaFeAsO, $J_{2}$ here is dominated by the superexchange bridged
by Se atoms \cite{Ma2008,Ma2008a}. According to the mechanism of
superexchange and the band structure of FeSe, we find that the superexchange
interaction is from the half-filled Fe d$_{xz}$/d$_{yz}$ orbitals
bridged by the Se p-orbitals, and Goodenough-Kanamori rules state
that this is AFM coupling \cite{Goodenough1955}. With the expanding
of the lattice constant, the out of plane height of Se atoms tends
to decrease and the Fe-Se-Fe angle (defined in Figure 4(b)) tends
to increase as shown in Table I. The increase of the Fe-Se-Fe angle
increases the overlapping of the Fe d$_{xz}$/d$_{yz}$ orbitals and
Se p-orbitals, while the Fe-Se-Fe superexchange interaction would
be maximized if the Fe-Se-Fe angle is $180\text{\textdegree}$ according
to Goodenough-Kanamori rules \cite{Goodenough1955}. From the above
discussions, we can conclude that the tensile strain enhances the
superexchange interaction and then enlarges the next-nearest-neighbor
coupling $J_{2}$, which directly stabilize the SDW state in monolayer
and bulk FeSe. 

\begin{figure}
\begin{centering}
\includegraphics[scale=1.0]{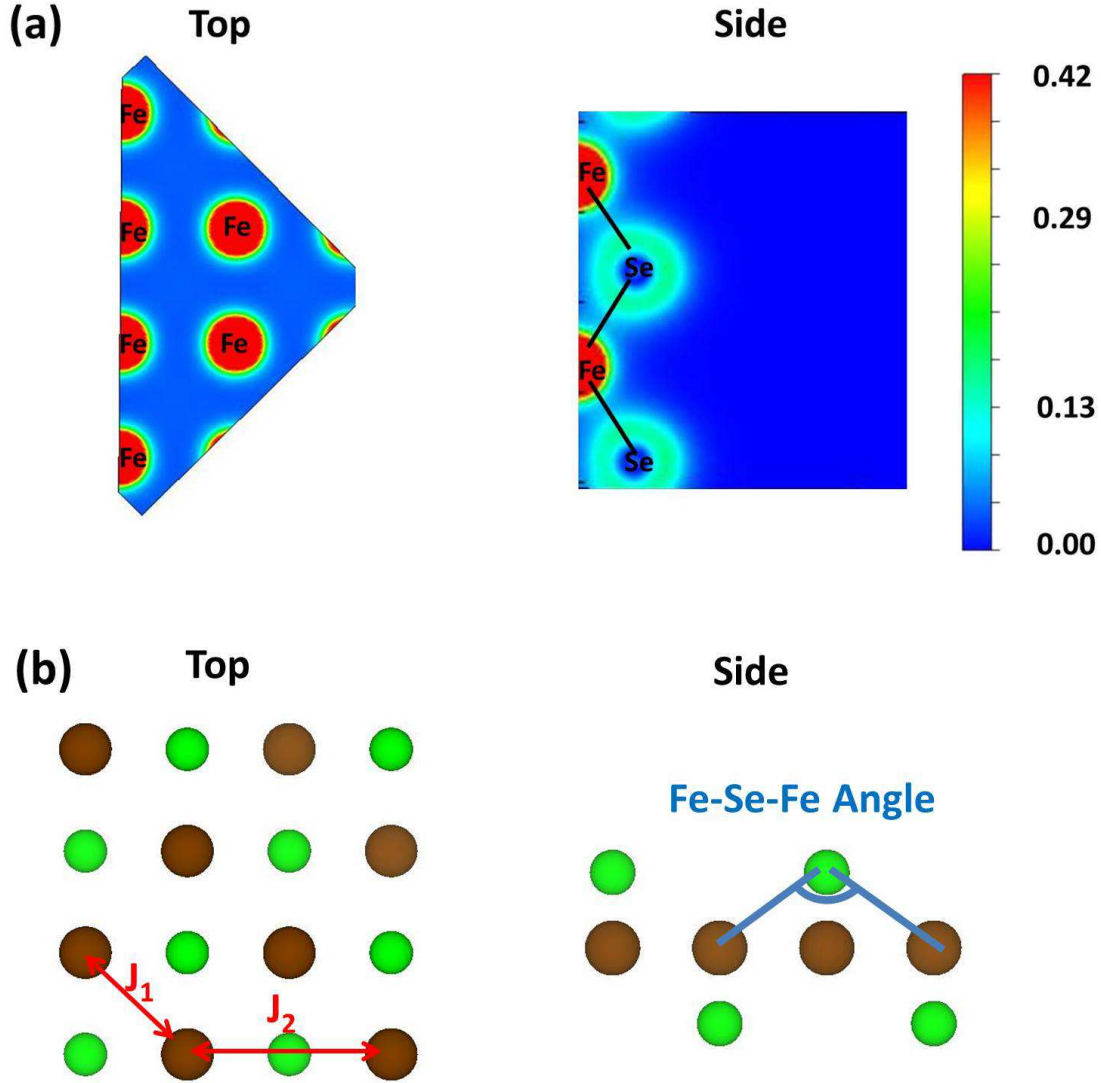}
\par\end{centering}

\centering{}\caption{(a) The top view and the side view of the charge distribution in monolayer
FeSe. In the top view, there is almost no bond between Fe atoms. In
the side view, we can clearly observe that there exists a bond (black
line) between Se and Fe. (b) The top view of the nearest as well as
the next-nearest magnetic exchange interacting $J_{1}$ and $J_{2}$
(red arrow) and the side view of the defined Fe-Se-Fe angle between
two next-nearest Fe atoms (blue line).}
\end{figure}

\begin{figure}
\includegraphics[scale=1.0]{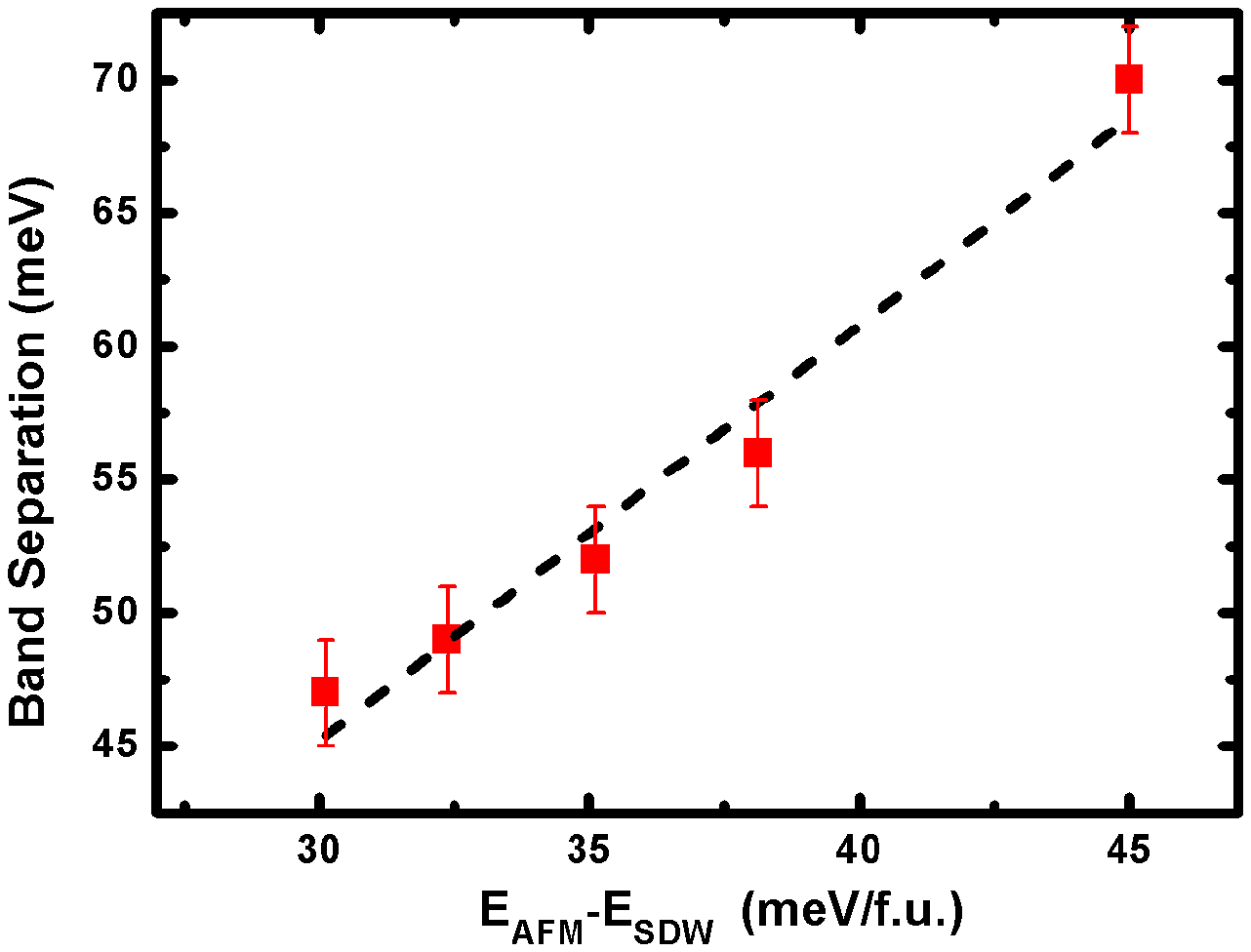}

\caption{The experimental band separations for FeSe films at various lattice
constants\cite{Tan2013} versus the calculated energy difference between
the AFM state and the SDW state. The dash line is a linear fitting.
The error bar for the experimental band separations is $\pm2$ meV. }
\end{figure}

\textcolor{black}{The recent experiment by Tan $et.$ $al.$ used
the band separation at M point to characterize the strength of the
SDW in FeSe thin films. The observed band separation increases with
expanding of the lattice constant, which is essntially caused by the
different band dispersions along the AFM and FM directions in the
SDW state \cite{Tan2013}. Here we observe a linear relation between
our calculated relative stability of the SDW and the experimental
band separation as shown in Figure 5. It indicates that our calculation
results of the energy difference is closely correlated to the experimental
observation, and it further implies that the band separation is a
relevant energy scale to characterize the strength of the SDW in FeSe
thin films.} This relation with the lattice constant is directly ascribed
to the enhanement of the next-nearest superexchange interaction due
to the increased Fe-Se-Fe angle with the lattice expansion. Recent
theoretical work predicts that a strong next-nearest AFM interaction
could lead to a high temperature superconductor \cite{Hu2012}. According
to this prediction, our reult shows that strain enhanced superexchange
interaction should play important role in the over 60 K high temperature
superconductivity observed in FeSe-STO.

\subsection{How The Interfacial Coupling Affects the SDW in FeSe}

For the monolayer FeSe on a defect-free STO substrate, we did not
find apparent charge transfer at the interface from Bader analysis
\cite{Tang2009}, and we did not yet find any obvious structural distortion
occured in monolayer FeSe. However, after detailed calculations of
the charge difference between the epitaxial and free-standing monolayer
FeSe, we find that the intefacial coupling would induce the redistribution
of the charge density in epitaxial monolayer FeSe, which would decrease
the charge density in the spin-majority d$_{xz}$/d$_{yz}$-orbital
states of Fe atoms as shown in the inset of Figure 6. If we gradually
increase the intereface distance from the equilibrium position 3.1
\AA \  to 5.6 \AA , the energy difference between the SDW state
and the AFM state rises from 37 meV/f.u. to 52 meV/f.u. as shown in
Figure 6. In the meantime, the charge redistribution occuring in monolayer
FeSe becomes smaller and smaller, which suggests that the charge redistribution
is attributed to the interfacial interaction. The decrease of the
charge density in the spin-majority d$_{xz}$/d$_{yz}$-orbital states
with decreasing interface distance would reduce the charge overlapping
between the Fe d$_{xz}$/d$_{yz}$-orbitals and the Se p-orbitals,
and then lower the superexchange interaction. As we have discussed
in Sec. III A, a smaller superexchange interaction would bring about
the weaker next-nearest-neighbor coupling J$_{2}$, and degrade the
stability of the SDW in FeSe. So the interfacial interaction between
monolayer FeSe and STO substrate could decrease the stability of the
SDW. 

\begin{figure}
\centering{}\includegraphics[scale=1.0]{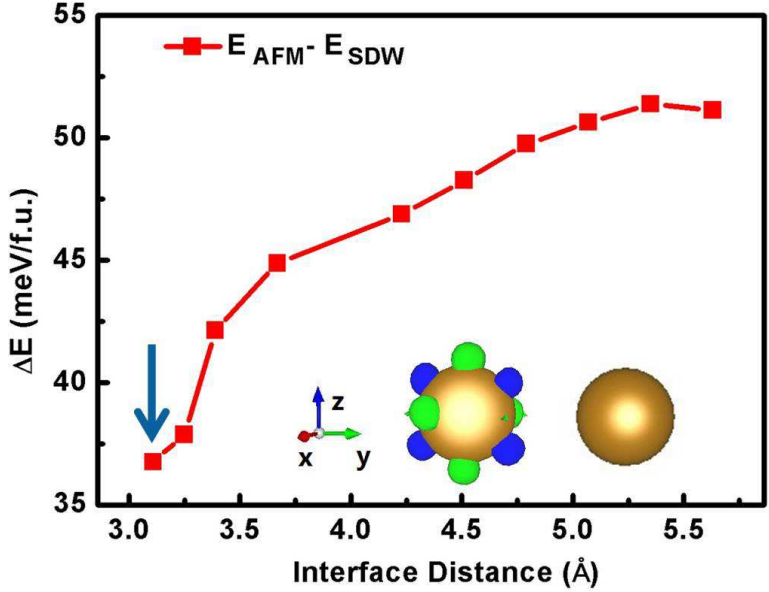}\caption{The calculated energy difference between the SDW state and the AFM
state as a function of the interface distance. With the decrease of
the interface distance, the SDW state becomese less stable relative
to the AFM state. The blue arrow shows the equilibrium distance as
3.1 \AA . Inset shows the charge redistribution on Fe atoms in epitaxial
FeSe at the equilibrium distance (left side). The blue part represents
the charge density lost while the green part represents the charge
density gained. The inset also shows there is almost no charge redistribution
on Fe atom if the interface distance is 5.6 \AA  \ (right side).}
\end{figure}

We find that the change of the d$_{xz}$/d$_{yz}$-orbital states
could be induced by the dipolar field along the direction perpendicular
to the interface. The plane-averaged charge density difference between
the monolayer FeSe-STO, free-standing monolayer FeSe and STO substrate
are shown in Figure 7(a). Although there is no charge transfer between
the monolayer FeSe and the STO substrate, the charge redistribution
near the interface can be observed clearly. It forms a charge dipole
at the interfacial region due to the interfacial coupling, which has
been previously explained by the metal-insulator band alignment \cite{Tersoff1984,Tersoff1985}.
The interfacial dipole formed at the monolayer FeSe-STO interface
could induce the electric field along the direction perpendicular
to the interface which induces the change of the d-orbital order of
Fe atom. 

\begin{figure}
\begin{centering}
\includegraphics[scale=1.0]{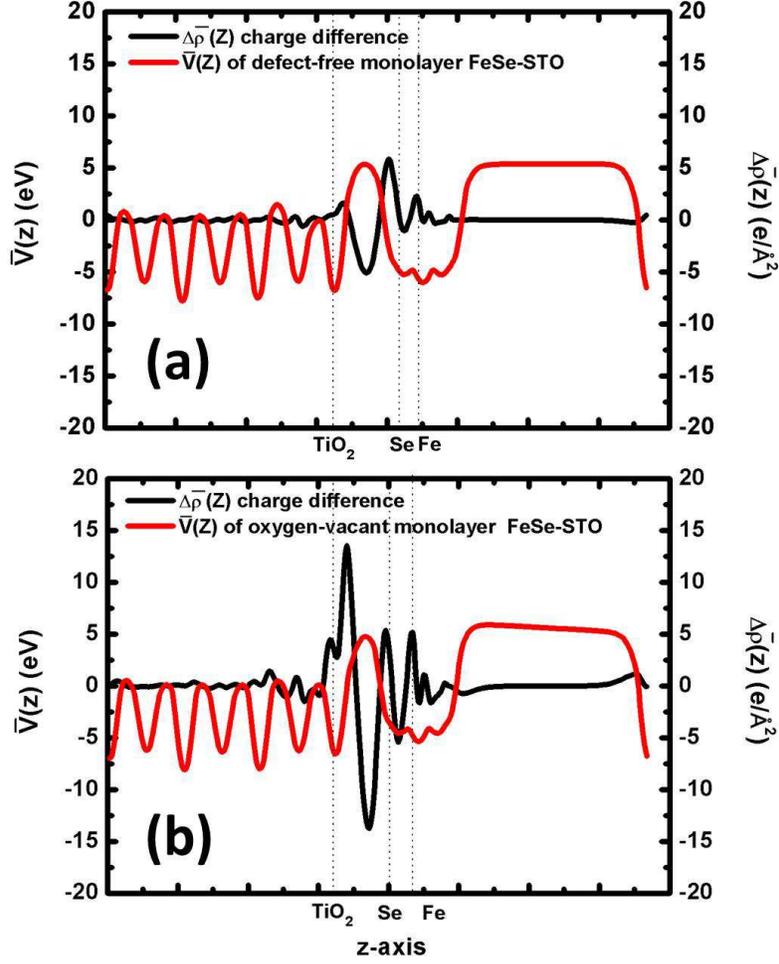}
\par\end{centering}

\caption{The plane-averaged charge density difference (black line) and the
local potential (red line) for (a) the monolayer FeSe/defect-free
STO interface and (b) the monolayer FeSe/oxygen-vacant STO interface.
The position of the surface TiO$_{2}$ layer of STO substrate and
FeSe atomic layers are denoted by the dotted vertical lines. Significant
charge transferred from the top TiO$_{2}$ layer of the oxygen-vacant
STO substrate to the Fe atomic layer can be observed here.}
\end{figure}

\subsection{How The Oxygen Vacancy Affects the SDW in FeSe}

Oxygen vacancy on the STO surface is experimentally unavoidable due
to the heat treatment in preparing the STO substrate \cite{Wang2012}.
In order to understand how the oxygen vacancy affects the property
of monolayer FeSe, we have simulated monolayer FeSe on STO substrate
with different concentration of oxygen vacancies.

Firstly, we calculate the energy difference between the NM state,
the AFM state and the SDW state, as well as the charge density distribution
in monolayer FeSe on the STO substrate with different concentration
of oxygen vacancies. The results are shown in Fig. 8. We find that
the oxygen vacancy can induce charge transfer from the STO substrate
to monolayer FeSe, the higher concentration of the oxygen vacancy,
the more the charge is transferred (Fig. 8(b)). As more and more charge
transferred to the monolayer FeSe, the stability of the SDW state
relative to the AFM state and the NM state decreases monotonously
(Fig. 8(a)). We also calculate the plane-averaged charge difference
obtained by substracting the valence charge densities of the free-standing
FeSe layer and isolated STO substrate from the monolayer FeSe on STO.
The results presented in Fig. 7(b) clearly show the monolayer FeSe
on STO substrate with oxygen vacancies inducing significant charge
transferred to Fe atoms while almost no charge transfer existing when
it is on the defect-free STO substrate as shown in Fig. 7(a). 

\begin{figure}
\begin{centering}
\includegraphics[scale=1.0]{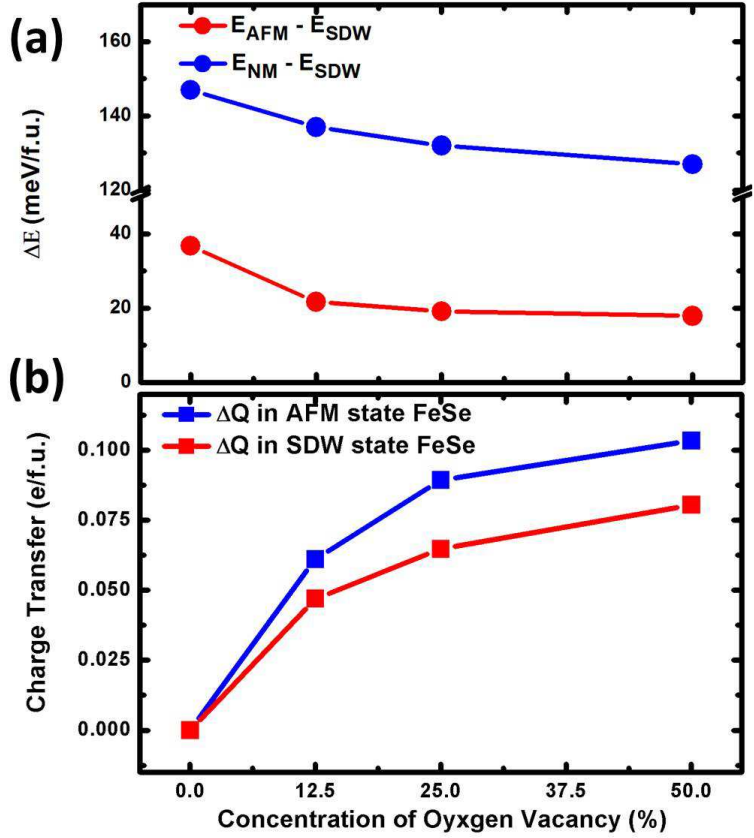}
\par\end{centering}

\caption{(a) The energy difference between the AFM state (blue circle), the
NM state (red circle) and the SDW state with different concentration
of oxygen vacancy. (b) The charge transferred to $2\times2$ supercell
of monolayer FeSe with the AFM state (blue square) and the SDW state
(red square) calculated by Bader analysis. The charge of monolayer
FeSe on defect-free STO substrate is set to be the reference. Both
the oxygen vacancy and the charge transferred from STO are related
to decrease the relative stability of the SDW state.}
\end{figure}

We find that the charge is transferred from the substrate to the spin-minority
d$_{xz}$/d$_{yz}$-orbital states of the Fe atom in monolayer FeSe,
so it would decrease the superexchange interaction and reduce the
stability of the SDW state. From Table II we find $J_{2}$ decreases
from 48 meV to 35 meV and $J_{2}/J_{1}$ decreases from 0.63 to 0.59
if 12.5\% oxygen vacancy exists on the surface of the STO substrate.
Based on the PDOS of Fe atom on STO substrate with 12.5\% oxygen vacancy,
we do find that about $\sim$0.1 charge transferred to the spin-minority
d$_{xz}$/d$_{yz}$-orbital states. Since the superexchange between
two next-nearest ideally half-field orbitals is the strongest, the
increasing charge density in the spin-minority d$_{xz}$/d$_{yz}$-orbital
states would decline the AFM superexchange coupling, and it would
decrease the stability of the SDW state. Moreover, we find that the
oxygen vacancy can magnetically polarize the nearby Ti atoms, and
the magnetism of Ti atoms could also decrease the stability of the
SDW state. For the concentration of 12.5\% oxygen vacancies, the nearby
Ti atom would possess a magnetic moment of 0.88 $\mu_{B}$ \cite{Park2013,Pavlenko2012a}.
The ferromagnetism from the Ti atoms would break the symmetry of spin
in monolayer FeSe, as shown in Figure 9. the PDOS of spin-minority
electrons of Fe atom would increase a lot near the Fermi level. 

\begin{table}
\begin{centering}
\begin{tabular}{|c|c|c|c|}
\hline 
 & J$_{1}$ (meV) & J$_{2}$ (meV) & J$_{2}$/J$_{1}$\tabularnewline
\hline 
\hline 
Free-standing Monolayer FeSe & 82 & 60 & 0.73\tabularnewline
\hline 
Monolayer FeSe on defect-free STO & 76 & 48 & 0.63\tabularnewline
\hline 
Monolayer FeSe on 12.5\% oxygen-vacant STO & 59  & 35 & 0.59\tabularnewline
\hline 
\end{tabular}
\par\end{centering}

\caption{Calculated nearest-neighbor and next-nearest-neighbor coupling parameters
of the free-standing monolayer FeSe, monolayer FeSe on defect-free
STO and monolayer FeSe on 12.5\% oxygen-vacant STO. Here we assume
S = 1 for Fe atoms.}
\end{table}

\begin{figure}
\begin{centering}
\includegraphics[scale=1.0]{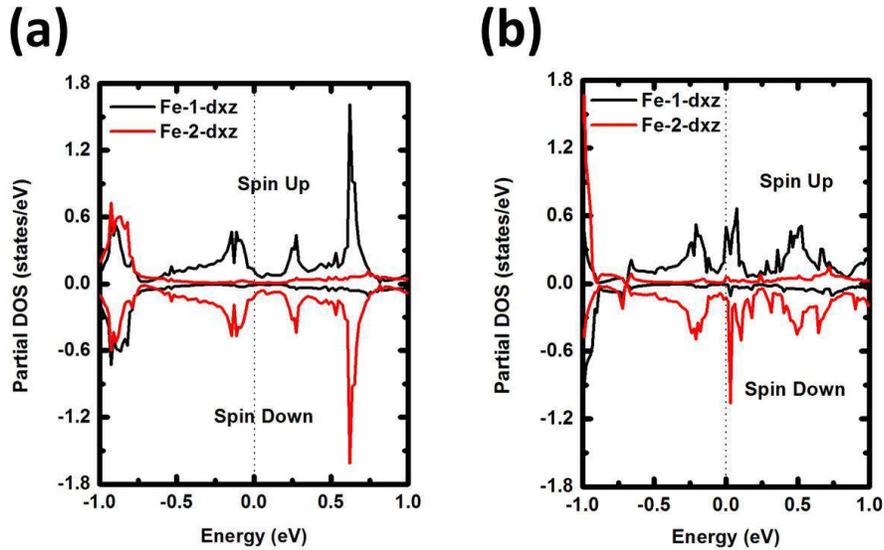}
\par\end{centering}

\centering{}\caption{The partial density of states (PDOS) of d$_{xz}$-orbital for two-type
Fe atoms in the unit cell of the SDW state with the up spin-majority
and down spin-majority electrons, repectively. The PDOS for Fe atoms
in expitaxial monolayer FeSe on (a) the defect-free STO substrate
and (b) the 12.5\% oxygen-vacant STO substrate. The existance of oxygen
vacancies on the surface of STO substrate disturb the symmetry of
spin of two-type Fe atoms and increases the density of states near
the Fermi energy significantly for spin-minority electrons of Fe atoms.}
\end{figure}

\section{summary }

We have studied the interfacial effect on the stability of the SDW
in monolayer FeSe using the GGA+U method. We find that tensile strain
can increase the superexchange interaction between the next-nearest
Fe atoms by increasing the Fe-Se-Fe bond angle, thus enhance the local
AFM coupling and the stability of the SDW. However, we also find that
the interfacial coupling between FeSe and STO substrate can change
the charge distribution in the 3d orbitals of Fe atoms, leading to
less charge density in the spin-majority d$_{xz}$/d$_{yz}$-orbitals.
It decreases the superexchange coupling between the next nearest Fe
atoms and suppress the stability of the SDW. In agreement with previous
calculation\cite{Bang2013}, we also observed a significant charge
transferred from the oxygen-vacant substrate to monolayer FeSe. Furthermore,
we find that such charge will be transferred to the spin-minority
d$_{xz}$/d$_{yz}$-orbitals of Fe atoms, and the almost fully occupied
d$_{xz}$/d$_{yz}$-orbitals will decline AFM superexchange interaction,
thus will also suppress the stability of the SDW. 

In summary, the substrate induced tensile strain can enhance the next-nearest
antiferromagnetic interaction, while the interfacial coupling and
charge transfer will destroy the long range magnetic order. We provide
a systematical microscopic description for the interfacial effects
on the magnetism and further suggest a strong correlation between
the magnetism and the possible high T$_{c}$ in monolayer FeSe on
STO substrate. Our results build the foundation for understanding
the prominent role of magnetism in this new kind of iron-based superconductor. 
\begin{acknowledgments}
HY Cao thanks Shiyou Chen for helpful discussions. The work was partially
supported by the Special Funds for Major State Basic Research, National
Natural Science Foundation of China (NSFC), Program for Professor
of Special Appointment (Eastern Scholar) and the National Basic Research
Program of China (973 Program). Computation was performed in the Supercomputer
Center of Fudan University.
\end{acknowledgments}
\pagebreak{}

\bibliographystyle{apsrev}
\bibliography{MyCollection}

\end{document}